\documentclass[dvipdfmx]{PoS}
\newcommand{\NS}{ N_{\rm S} }
\newcommand{\NT}{ N_{\rm T} }
\newcommand{\NF}{ N_{\rm F} }
\newcommand{\ml}{ m_{\rm l} }
\newcommand{\ms}{ m_{\rm s} }

\newcommand{\kl}{ \kappa_{\rm l} }
\newcommand{\ks}{ \kappa_{\rm s} }

\usepackage{amsmath}

\title{Critical endpoint in the continuum limit and critical endline at $\NT=6$ of the finite temperature phase transition of QCD with clover fermions}

\ShortTitle{Phase diagram of the finite temperature phase transition of QCD}

\author{\speaker{Yoshifumi Nakamura}\\
        RIKEN Center for Computational Science, Kobe, Hyogo 650-0047, Japan\\
        Graduate School of System Informatics, Department of Computational Sciences, Kobe University, Kobe, Hyogo 657-8501, Japan\\
        E-mail: \email{nakamura@riken.jp}}

\author{Yoshinobu Kuramashi\\
        Faculty of Pure and Applied Sciences, University of Tsukuba, Tsukuba, Ibaraki 305-8571, Japan\\
        Center for Computational Sciences, University of Tsukuba, Tsukuba, Ibaraki 305-8577, Japan\\
        E-mail: \email{kuramasi@het.ph.tsukuba.ac.jp}}

\author{Hiroshi Ohno\\
        Center for Computational Sciences, University of Tsukuba, Tsukuba, Ibaraki 305-8577, Japan\\
        E-mail: \email{hohno@ccs.tsukuba.ac.jp}}

\author{Shinji Takeda\\
        Institute of Physics, Kanazawa University, Kanazawa 920-1192, Japan\\
        E-mail: \email{takeda@hep.s.kanazawa-u.ac.jp}}

\abstract{We investigate the critical endpoints of the finite temperature
phase transition of QCD at zero chemical potential.
We employ the renormalization-group improved Iwasaki gauge action
and non-perturbatively O(a)-improved Wilson-clover fermion action.
The critical endpoints are determined by using the intersection
point of kurtosis, employing the multi-parameter,
multi-ensemble reweighting method.
We present results for the critical endline at $\NT$ = 6 and the
continuum extrapolation for the critical endpoint of
the SU(3)-flavor symmetric point.}

\FullConference{37th International Symposium on Lattice Field Theory - Lattice2019\\
		16-22 June 2019\\
		Wuhan, China}

\begin{document}
%----------------------------------------------------------------------------
\section{Introduction}\label{intro}
The nature of the finite temperature phase transition
of 2+1 flavor QCD at zero chemical potential depends on quark masses.
The order of transition and universality class  are summarized in the plane of light quark mass, $\ml$ and strange quark mass, $\ms$,
which is called the Columbia plot~\cite{Brown:1990ev}.
%\begin{figure}[thb]
%  \centering
%  %\includegraphics[bb= 0 0 695 634,width=4cm]{./figure/columbia.png}
%  \includegraphics[width=5cm]{./figure/columbia.png}
%  \caption{Columbia plot.}
%  \label{fig-Columbia}
%\end{figure}
%
%Figure~\ref{fig-Columbia} shows current understanding  of  Columbia plot.

A first order phase transition is expected in the small quark mass region~\cite{Pisarski:1983ms}.
Many lattice QCD studies have shown that the phase transition is also of first order in the heavy quark mass region
while it is crossover in the medium quark mass region.
The boundary between the first order and crossover regions is a second 
order phase transition of $Z_2$ universality class.

The nature in the lower-left corner of the Columbia plot has not been fully understood yet.
The first lattice QCD calculation was done by using standard Wilson fermions at $\NT=4$ roughly 20 years ago.
It reported the critical mass at the critical endpoint (CEP), $m_{\rm E}$, for $\NF=3$ is heavy: 
the critical quark mass $m_{\rm q, E} = m_{\rm l, E} = m_{\rm s, E}\gtrsim 140$ MeV or, equivalently, the critical pseudo scalar mass $m_{\rm PS, E} = m_{\pi, {\rm E}} = m_{\eta_s, {\rm E}}\gtrsim 1$ GeV
~\cite{Iwasaki:1996zt}.
After a preliminary study with standard Wilson gauge and staggered fermions which reported the bare critical mass $am_{\rm q, E} \sim 0.035$~\cite{Aoki:1998gia} at $\NT=4$,
%https://doi.org/10.1016/S0920-5632(99)85104-4
Karsch {\it et al.} reported preliminary values for the critical mass,
$m_{\rm PS, E} \sim 290$ MeV with unimproved gauge and staggered fermion actions and 
$m_{\rm PS, E} \sim 190$ MeV with improved gauge and p4 staggered fermion actions~\cite{Karsch:2001nf}.
These results were obtained by using the R-algorithm~\cite{Gottlieb:1987mq}.
Afterward, the results were updated as 
$m_{\rm PS, E} = 290(20)$ MeV with unimproved gauge and staggered fermion actions and 
$m_{\rm PS, E} =   67(17)$ MeV with improved gauge and p4 staggered fermion actions~\cite{Karsch:2003va}.
Then, in ref.~\cite{deForcrand:2006pv}, de Forcrand and Philipsen  obtained $am_{\rm q, E} = 0.0260(5)$ by using the RHMC algorithm~\cite{Clark:2004cp,Clark:2006fx}, which is about 25\% smaller than 
the value $am_{\rm q, E} \approx 0.033$ quoted by works using the R-algorithm.
They also performed $\NF=2+1$ simulations and obtained the critical line
and tri-critical point, $a\ms^{\rm tri} \approx 0.7$, where lattice spacing $a$ was approximately $0.3$ fm.
In ref.~\cite{deForcrand:2007rq} with unimproved staggered fermions, it was reported that the ratio of $m_{\rm PS, E}$ and the CEP temperature $T_{\rm E}$
decreased from  $1.680(4)$ to $0.954(12)$ as increasing $\NT$ from $4$ to $6$.
These results are showing very large cut off effect for the critical mass and it is important to increase $\NT$ and use improved actions.
Further studies with improved staggered fermions have not found the first order phase transition and quoted only 
a bound of the critical mass, $m_{\rm PS, E} \lesssim   50$ MeV,~\cite{Endrodi:2007gc, Ding:2011du, Bazavov:2017xul}.
Therefure, the positions of the critical endline (CEL), $\ms^{\rm tri}$ and CEP for $\NF=3$
are still particularly important problems to be solved at this moment.

Recently we also have investigated the nature of the finite phase transition in the small quark mass region
by using non-perturbatively $O(a)$-improved Wilson-clover fermions.
We have determined CEP at $\NT=4, 6, 8$, and $10$ as well as an upper bound of 
CEP in the continuum limit for $\NF=3$~\cite{Jin:2014hea, Jin:2017jjp}.
For $\NF=2+1$, we have studied at $\NT=6$ and determined CEL around  the SU(3) flavor symmetric point. 
Then, we confirmed that the slope of CEL at the SU(3) flavor symmetric point is -2~\cite{Kuramashi:2016kpb}.
In this paper, we extend our study for both CEP at the SU(3) flavor symmetric point and 
CEL away from the SU(3) flavor symmetric point.

For $\NF=4$, comparison between Wilson and staggered fermion, in no rooting concern for staggered fermion setup, is also ongoing~\cite{1702.00330, 1812.01318}.
%----------------------------------------------------------------------------
\section{Simulations}\label{simu}

We employ the renormalization-group improved Iwasaki gauge action~\cite{Iwasaki:2011np} and
non-perturbatively $O(a)$-improved Wilson-clover fermion action~\cite{Aoki:2005et}.
CEP is determined by using the intersection point of kurtosis of chiral condensate.
This method is expounded in Ref~\cite{Jin:2014hea} and used in our recent studies~\cite{Jin:2014hea, Jin:2017jjp, Kuramashi:2016kpb}.
Expectation value, susceptibility and skewness of chiral condensate are also used for confirming phase transition and
determination of the transition point.
Chiral condensate and its higher moments are computed from traces of the inverse Wilson clover Dirac operator up to 
a power of $-4$, {\it i.e.} ${\rm Tr} D^{-1, -2. -3, -4}$, by using 10 noise vectors. We have checked that 10 noises are good enough 
in this study.
We employ the multi-parameter, multi-ensemble reweighting method~\cite{Ferrenberg:1988yz} to determine CEP 
with very small statistical error.
We reweight both $\kl$ and $\ks$ so that we can determine many CEPs without doing simulations at many parameter sets.
We performed zero temperature runs for physical scale setting which are covering almost all transition points of finite temperature simulations.
Lattice spacings are computed by the Wilson flow lattice scale $\sqrt{t_0}/a$ ~\cite{Luscher:2010iy}. 
Our finite temperature simulations are performed at the temporal size $\NT=6$ and with a lattice spacing $a \approx$0.19 fm for CEL, and
$\NT=12$ and $a \approx$ 0.12 fm for the continuum limit of CEP at the SU(3) flavor symmetic point. 
The spatial size $\NS$ is  $10, 12, 16$, and $24$ at $\NT=6$ and $\NS=16,20,24,28$, and $32$ at $\NT=12$.
We have confirmed $m_{\rm PS}L > 4$ at almost all transition points, where $m_{\rm PS}$ is the pseudo scalar mass
and $L$ is the physical spatial extent.
We will explain complete simulation details in our upcoming full paper.

\section{Simulation results}\label{resu}
\subsection{Simulation results for $\NF=3$}\label{resu3}

\begin{figure}[thb]
  \centering
  \includegraphics[width=0.24\textwidth]{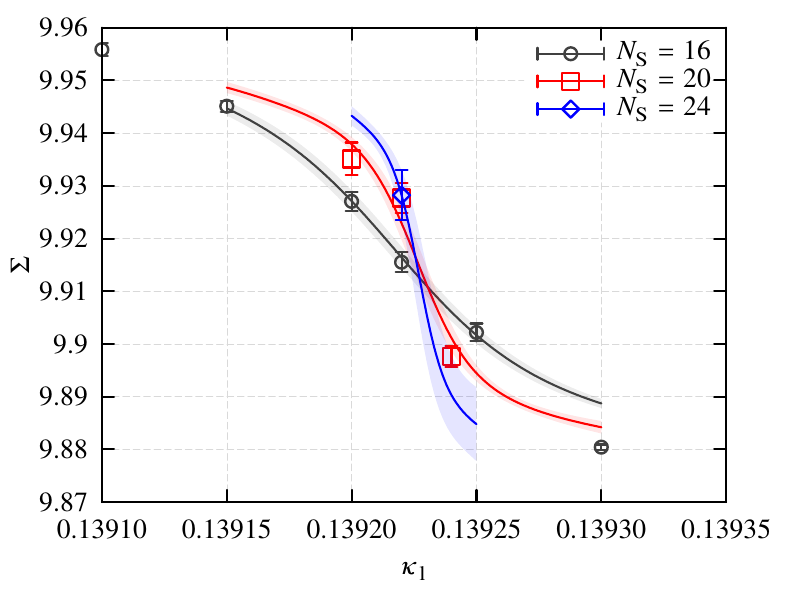}
  \includegraphics[width=0.24\textwidth]{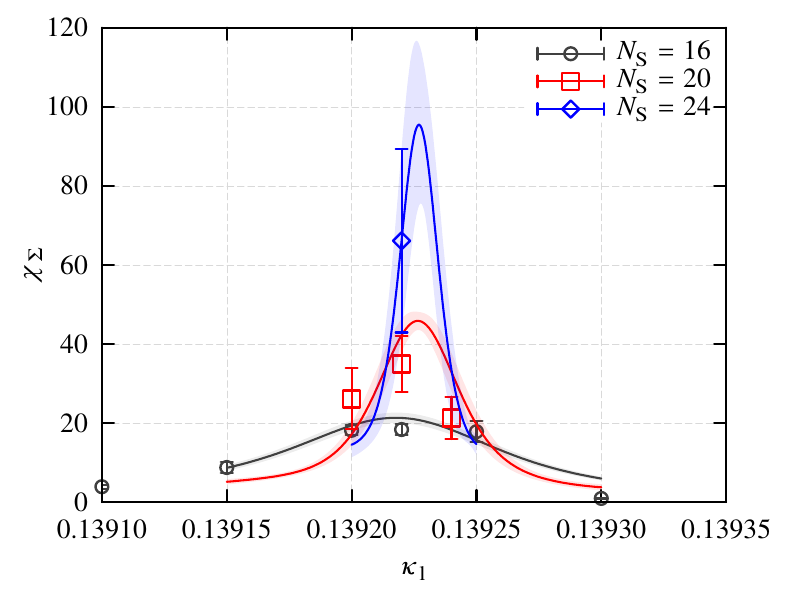}
  \includegraphics[width=0.24\textwidth]{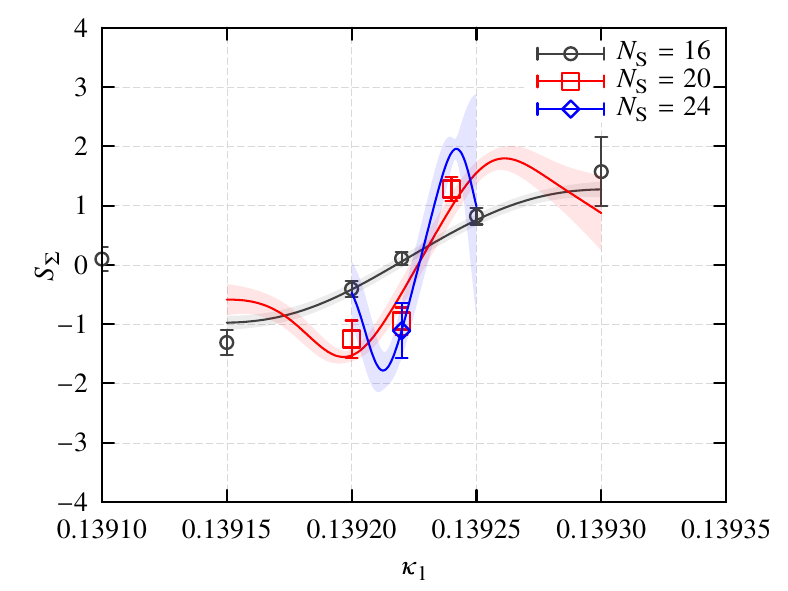}
  \includegraphics[width=0.24\textwidth]{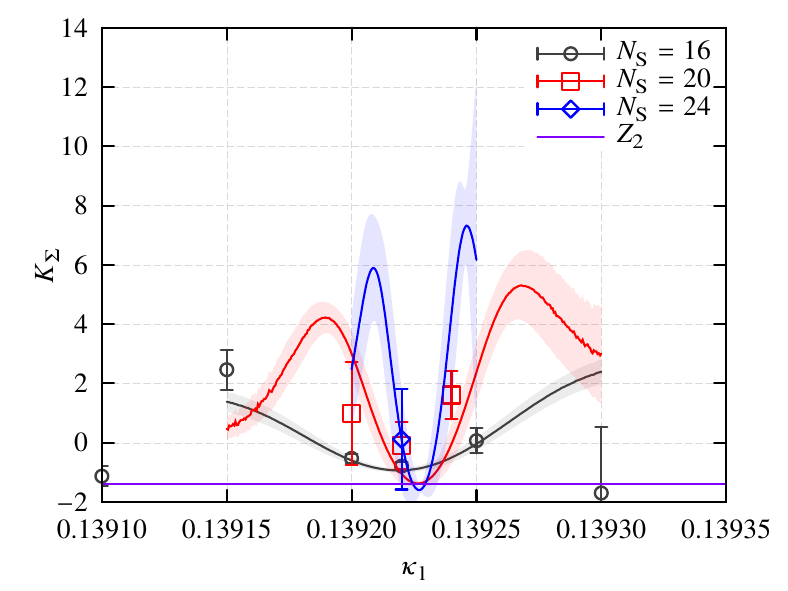}
  %\vspace{5mm}
  \caption{Expectation value , susceptibility, skewness and kurtosis of chiral condensate as a function of $\kappa$ at $\beta = 1.80$.}
  \label{fig-evsk}
\end{figure}

\begin{figure}[thb]
  \centering
  \includegraphics[width=0.35\textwidth]{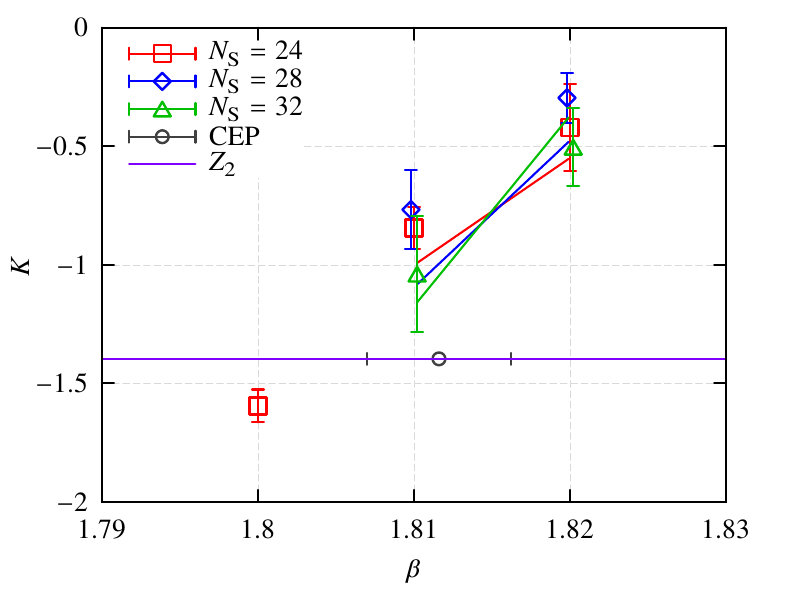}
  \includegraphics[width=0.35\textwidth]{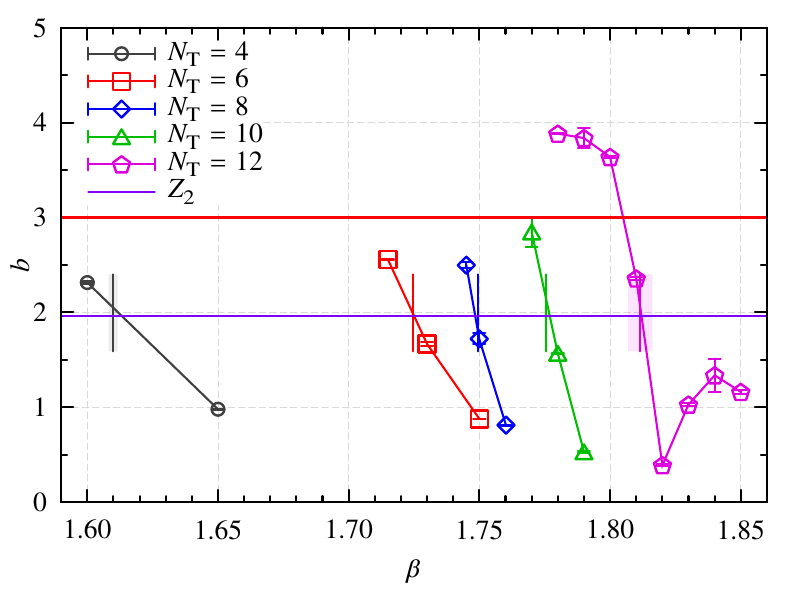}
  %\vspace{5mm}
  \caption{Kurtosis intersection (left) and the ratio of the critical exponents, $b=\gamma/\nu$ (right) as a function of $\beta$ for $\NF=3$
  including smaller $\NT$ results.}
  \label{fig-intersect}
\end{figure}

We show expectation value, susceptibility, skewness and kurtosis of chiral condensate at $\beta = 1.80$ as example in Fig.~\ref{fig-evsk}.
It shows that the reweighting method works well and we can find the phase transition precisely.
Fig.~\ref{fig-intersect} shows a kurtosis intersection plot and a plot for the ratio of the critical exponents, $b=\gamma/\nu$ determined by 
finite size scaling of the peak height of susceptibility, $\chi_{\rm max} = \NS^b$.
To locate the intersection point of kurtosis we use a following modified fitting form including a correction term
from energy-like observable that we have used in our previous study:
\begin{equation}
K = \left[ K_{\rm E} + A\NS^{1/\nu}(\beta-\beta_{\rm E}) \right] (1+B\NS^{y_{\rm t}-y_{\rm h}})\,,
\end{equation}
wherer $K$, $K_{\rm E}$, $\beta_{\rm E}$, $y_{\rm t}$, and $y_{\rm h}$ are kurtosis, kurtosis at the endpoint,
$\beta$ at the endpoint, the exponent for the temperature and the magnetic field, respectively.
We examine three fits as follows. 
Fit-1 has no correction term ($B=0$) and all other parameters are used as fit parameters. 
Fit-2 also neglects the correction term assuming the 3D $Z_2$ universality class for $K_{\rm E}$ and $\nu$. 
Fit-3 includes the correction term assuming the 3D $Z_2$ universality class for $K_{\rm E}$, $\nu$, and $y_{\rm t}-y_{\rm h}$.
In the 3D $Z_2$ universality class, $K_{\rm E} = -1.396$, $\nu=0.63$, and $y_{\rm t}-y_{\rm h} = -0.894$.
The fit results are summarized in Table~\ref{tab-1}. 
Fit-1 gives substantially larger $K_{\rm E}$ than the 3D $Z_2$ value with large error. 
$\nu$ is consistent with the universal value $0.63$ but it has fairly large error. 
So we can not confirm transition belongs to the 3D $Z_2$ universality class from Fit-1.
We observe $\chi^2/{\rm d.o.f.}$ of Fit-2, assuming the 3D $Z_2$ universality class without correction term, is not bad
and reasonable $\chi^2/{\rm d.o.f.}<1$ for Fit-3.

For a cross check of the endpoint location, a $b$ plot (right panel of fig.~\ref{fig-intersect}) is helpful
because $b$ changes from dimension number ($3$ in this study) to $0$ via a certain value at the critical endpoint
when the transition changes from the first order phase transtion to crossover.
The value of a green horizontal line is $b$ of the 3D $Z_2$ universality class. We see $b$ as a fuction of $\beta$ is 
crossing the green line at $\beta_{\rm E}$ determined by the kurtosis intersection. This cross check tells our analysis works well.
We adopt $\beta_{\rm E}$ determined by Fit-3 in the following analysis.
Since $b$ of other universality classes is a similar value as $b$ of the 3D $Z_2$ universality class,
this plot is not suited to distinguish the universality class.

\begin{figure}[thb]
  \centering
  \includegraphics[width=0.35\textwidth]{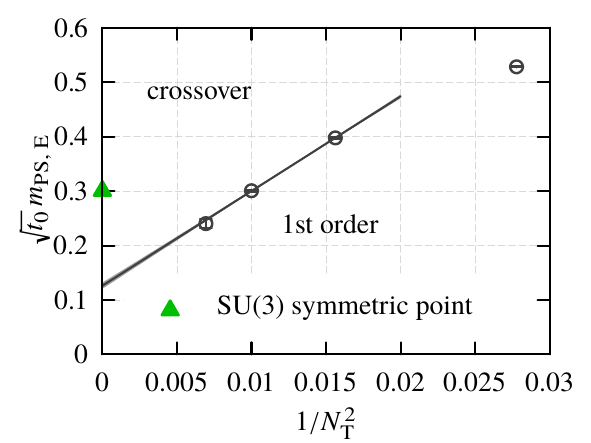}
  \includegraphics[width=0.35\textwidth]{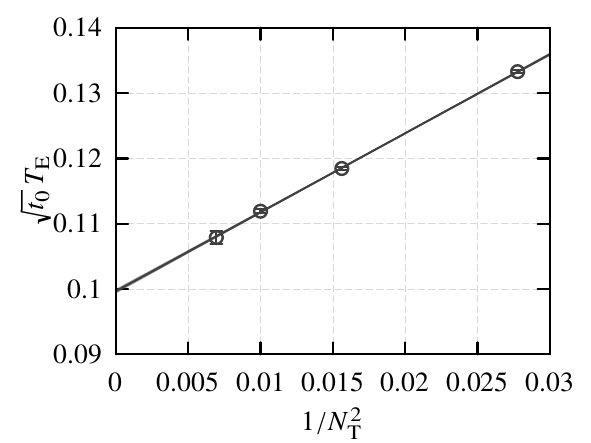}
  \caption{Continuum extrapolations for $\sqrt{t_0}m_{\rm PS, E}$ (left) and $\sqrt{t_0}T_{\rm E}$ (right).}
  \label{fig-nf3}
\end{figure}

In Fig.~\ref{fig-nf3},  $m_{\rm PS, E}$ and $T_{\rm E}$ normalized by $\sqrt{t_0}$ as a function of $1/\NT^2$ are shown.
Linear continuum extrapolations give 
$\sqrt{t_0}m_{\rm PS} =  0.1262(57)$ with $\chi^2/{\rm d.o.f.}=0.84$ and $\sqrt{t_0}T = 0.09968(36)$ with  $\chi^2/{\rm d.o.f.}=0.34$

\begin{table}[thb]
  \centering
  \caption{Kurtosis intersection fitting reuslts for $\NF=3$.}
  \label{tab-1}% Give a unique label
  \begin{tabular}{rllllllrr}\hline
  Fit&$\beta_{\rm E}$&$K_{\rm E}$&$\nu$&$A$&$B$&$y_{\rm t}-y_{\rm h}$&$\chi^2/{\rm d.o.f.}$\\\hline 
  $1$&$1.8145 ( 42 )$&$-0.64 ( 21 )$&$0.66 ( 56 )$&$0.4 ( 1.5 )$&$0$&$0$&$0.34$\\
  $2$&$1.7954 ( 66 )$&$-1.396$&$0.63$&$0.211 ( 69 )$&$0$&$0$&$1.34$\\
  $3$&$1.8098 ( 26 )$&$-1.396$&$0.63$&$0.419 ( 89 )$&$-7.0 ( 1.5 )$&$-0.894$&$0.29$\\\hline
  \end{tabular}
\end{table}

%----------------------------------------------------------------------------
\subsection{Simulation results for $\NF=2+1$}\label{resu2p1}

\begin{figure}[thb]
  \centering
  \includegraphics[width=0.32\textwidth]{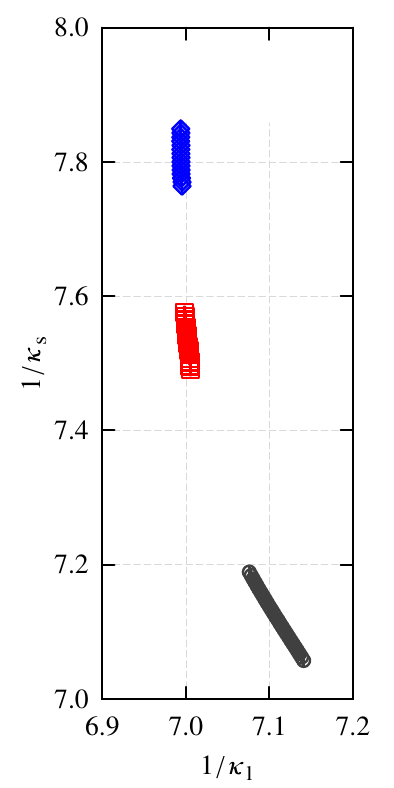}
  \includegraphics[width=0.32\textwidth]{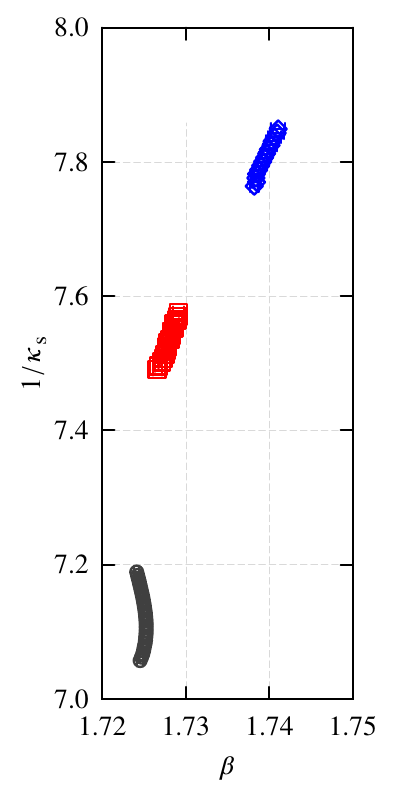}
  \caption{CEP in bare parameter planes.}
  \label{fig-ep}
\end{figure}

\begin{figure}[thb]
  \centering
  \includegraphics[width=0.32\textwidth]{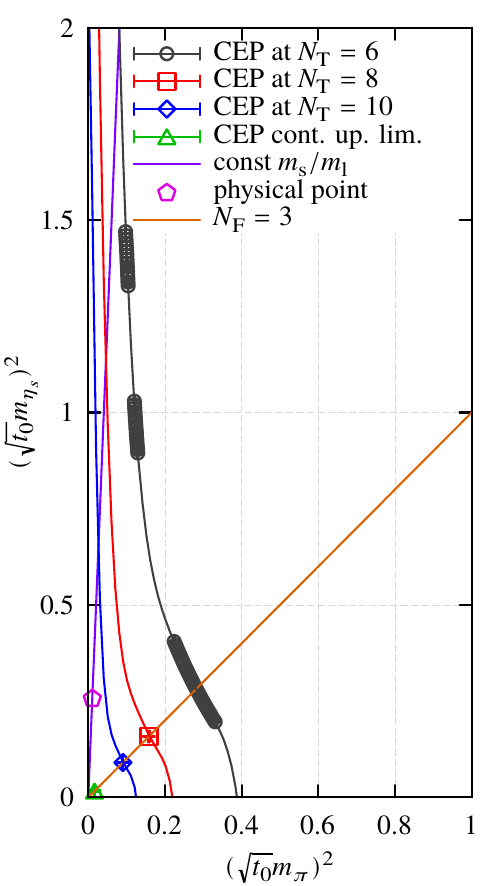}
  \includegraphics[width=0.32\textwidth]{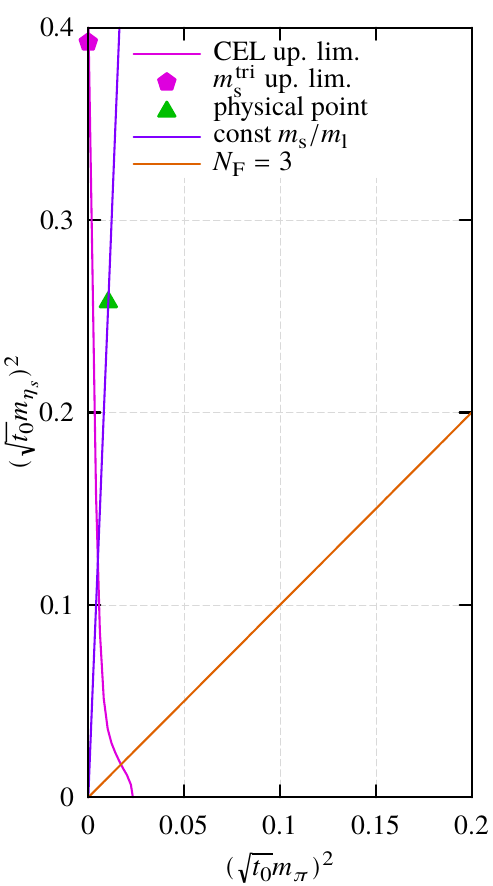}
  \caption{Preliminary results for CEL at $\NT=6, 8$, and $10$ (left), and in the continuum limit (right).}
  \label{fig-el}
\end{figure}

In Fig.~\ref{fig-ep}, we plot CEP in two different bare parameter planes: ($1/\kl$, $1/\ks$)-plane and ($\beta$, $1/\ks$)-plane. 
Fig.~\ref{fig-el} shows CEP at $\NT=6$ together with preliminary results for CEL at $\NT=6, 8$, and $10$ as well as 
its continuum extrapolation in a dimensionless physical scale plane. 
This plane corresponds a light quark and strange quark mass plane. 
We see that CEL blows up rapidly as decreasing the light quark mass.
We will explain how to estimate CEL at $\NT> 6$ later.

To estimate CEL, we first fit CEP at $\NT=6$ to a formula inspired by the tri-critical scaling law~\cite{Rajagopal:1995bc}
\begin{equation}\label{fit1}
y=b_0 + b_1 x^{2/5}\,,
\end{equation}
where $x=(\sqrt{t_0}m_{{\pi},{\rm E}})^2$, $y=(\sqrt{t_0}m_{{\eta_s},{\rm E}})^2$, $b_0$ and $b_1$ are free parameters.
Resulting $b_0$ is proportional to the strange quark mass at the tri-critical point.
% up to first order in chiral symmetry breaking.
We obtain $b_0=6.71(8)$ and $b_1=-13.3(2)$ with $\chi^2/{\rm d.o.f.}=0.54$ by using only data in range $x<0.105$, {\it i.e.} $m_{{\eta_s},{\rm E}}^2 / m_{{\pi},{\rm E}}^2 > 10$.
By changing the fitting range in this fitting, we can discuss the tri-critical scaling region.
Since the fitting which includes data up to $x \approx 0.125$ gives reasonable $\chi^2/{\rm d.o.f.}$, that is less than one, 
we update the tentative scaling region at $\NT=6$, {\it i.e.} $m_{{\eta_s},{\rm E}}^2 / m_{{\pi},{\rm E}}^2 > 7.5$.

In the all range fitting we use the formula adding a power series up to terms of order $x^5$ to eq.~(\ref{fit1}) 
\begin{equation}
y=b_0 + a_0 x^{2/5} + \sum_{i=1}^5 a_i x^i \,,
\end{equation}
where $b_0=6.71$ and $a_i$ are free parameters. By fixing $b_0$ we obtain reasonable $\chi^2/{\rm d.o.f.}$, $1.3$.
We could not find reasonable fitting neither without fixing $b_0$ nor with less polynomial order fitting functions.

%This value is roughly four times smaller than the ratio of the physical strange and light quark mass.

Further estimation is possible by using results for $\NF=3$ and assuming that 
there is no $\NT$ dependence in the shape of CEL.
The normalized pseudo scalar masses at CEP and the SU(3) flavor symmetric point
for each $\NT$, $\sqrt{t_0}m^{\rm sym}_{{\rm PS, E}, \NT}$ are

\begin{equation}
\begin{split}
\sqrt{t_0}m^{\rm sym}_{\rm PS, E, 6} &=0.5282(12)\,, \quad\sqrt{t_0}m^{\rm sym}_{\rm PS, E, 8} =0.3977(19)\,,\\
\sqrt{t_0}m^{\rm sym}_{\rm PS, E, 10} &=0.3006(19)\,, \quad \sqrt{t_0}m^{\rm sym}_{\rm PS, E, \infty} <0.1281(61)\,,
\end{split}
\end{equation}
where the result in the continuum limit (at $\NT=\infty$) is an upper bound.
Since the updated continuum limit of CEP at the SU(3) flavor symmetric point is still preliminary,
we estimate the upper bound of CEL in the continuum limit by using previous published results~\cite{Jin:2017jjp}.
For example, we obtain CEL at $\NT=8$ by scaling $\sqrt{t_0}m_{\pi,{\rm E}}$ and $\sqrt{t_0}m_{{\eta_s},{\rm E}}$
by the ratio of $\sqrt{t_0}m^{\rm sym}_{\rm PS, E, 8}$ to $\sqrt{t_0}m^{\rm sym}_{\rm PS, E, 6}$.
%
%Note that, we are ignoring lattice artifact in the estimation of CEL except for $\NT=6$ and
%CEL in the continuum limit of the right plot of Fig.~\ref{fig-el} is also the upper bound
%since we have only determined the upper bound of $\sqrt{t_0}m^{\rm sym}_{\rm PS, E}$ in the continuum limit.
%With these caveat, we find $\ms^{\rm tri} \lesssim 1.5 \, \ms^{\rm phy}$.
On the above assumption, we find $\ms^{\rm tri} \lesssim 1.5 \, \ms^{\rm phy}$ in the continuum limit.

%%%%%%%%%%%%%%%%%%%%%%%%%%%%%%%%%%%%%%%%%%%%%%%%%%%%%%%%%%%%%%%%%%%%%%%%%%%%%%%%%%%%%%%%%%%%%%%%%%%
\section{Summary}\label{summ}
We have determined  CEP at the SU(3)-flavor symmetric point at $\NT=12$, and 
CEL away from the SU(3)-flavor symmetric point
at $\NT=6$ with non-perturbatively $O(a)$-improved Wilson fermions.
We presented preliminary results for CEP in the continuum limit and
CEL at $\NT=8, 10$ and in the continuum limit.
We found that a linear continuum extrapolation including new data is reasonable and gives 
preliminary results at CEP in the continuum limit:
$\sqrt{t_0}m_{\rm PS} =  0.1262(57)$, $\sqrt{t_0}T = 0.09968(36)$.
Moreover, 3 series of multi-ensemble, multi-parameter reweighting well determine CEL,
where CEL at $\NT=6$ is nice agreement with $\ms -\ms^{\rm tri} \sim \ml^{2/5}$ in the small $\ml$ region,
with $\ms^{\rm tri} \lesssim 1.5 \, \ms^{\rm phy} $ as a very preliminary result.

%%%%%%%%%%%%%%%%%%%%%%%%%%%%%%%%%%%%%%%%%%%%%%%%%%%%%%%%%%%%%%%%%%%%%%%%%%%%%%%%%%%%%%%%%%%%%%%%%%%
%\section{Acknowledgements}
This research was supported by Multidisciplinary Cooperative Research Program in CCS, University of Tsukuba
and projects of the RIKEN Supercomputer System.
%This work is supported by JSPS KAKENHI Grant Numbers 26800130,
%FOCUS Establishing Supercomputing Center of Excellence.

\end{document}